\documentclass[10pt,a4paper]{article}

\usepackage{amssymb}
\usepackage{amsfonts}
\usepackage{mathrsfs}
\usepackage{latexsym}
\usepackage{amsmath}
\usepackage{amsthm}
\usepackage{amscd}               
\usepackage{graphics}
\usepackage{graphicx}
\usepackage{hhline}                   
\usepackage{euscript}               
\usepackage{rotating}
\usepackage{setspace}                 
\usepackage{fancyheadings}            
\usepackage[dvips]{hyperref}
\DeclareSymbolFont{NumBold}{U}{bbold}{m}{n}
\DeclareSymbolFontAlphabet{\mathNumbb}{NumBold}
\DeclareMathSymbol{\Id}{\mathord}{NumBold}{`1}
\newcommand{\mbf}[1]{\mbox{\boldmath{$#1$}}}

\newcommand{\mr}{\mathrm}

\newcommand{\D}{\mathcal{D}}
\newcommand{\E}{\mathcal{E}}
\newcommand{\F}{\mathcal{F}}

\renewcommand{\L}{\mathcal{L}}

\newcommand{\N}{\mathcal{N}}

\renewcommand{\P}{\mathcal{P}}

\renewcommand{\S}{\mathcal{S}}

\newcommand{\V}{\mathcal{V}}
\newcommand{\W}{\mathcal{W}}

\renewcommand{\a}{\alpha}
\renewcommand{\b}{\beta}
\newcommand{\g}{\gamma}
\renewcommand{\d}{\delta}
\newcommand{\eps}{\varepsilon}

\renewcommand{\r}{\rho}

\newcommand{\vf}{\varphi}
\renewcommand{\l}{\lambda}
\newcommand{\s}{\sigma}
\renewcommand{\t}{\tau}

\newcommand{\gG}{\Gamma}
\newcommand{\gD}{\Delta}
\newcommand{\gL}{\Lambda}

\newcommand{\e}{\mathop{\mathrm{e}}\nolimits}

\newcommand{\Tr}{\mathop{\mathrm{Tr}}\nolimits}

\newcommand{\Ch}{\mathop{\mathrm{Ch}}\nolimits}
\newcommand{\Td}{\mathop{\mathrm{Td}}\nolimits}
\newcommand{\Ind}{\mathrm{Ind}}
\newcommand{\supp}{\mathop{\mathrm{supp}}\nolimits}

\newcommand{\dd}{\mathrm{d}}
\newcommand{\pd}{\partial}

\newcommand{\Real}{\mathbb{R}}
\newcommand{\Compl}{\mathbb{C}}
\newcommand{\Integer}{\mathbb{Z}}
\newcommand{\Sphere}{\mathbb{S}}

\let\IM=\Im
\let\RE=\Re
\let\Im=\undefined
\let\Re=\undefined
\newcommand{\Im}{\mathop{\IM\mathfrak{m}}\nolimits}
\newcommand{\Re}{\mathop{\RE\mathfrak{e}}\nolimits}

\newcommand{\ds}{\displaystyle}
\newcommand{\cc}{\overline}

\newcommand{\tld}{\widetilde}
\newcommand{\isom}{\simeq}

\newcommand{\<}{\langle}
\renewcommand{\>}{\rangle}

\newcommand{\res}{\mathop{\mathrm{res}}}

\renewcommand{\k}{\mbf{k}}

\newcommand{\SU}{\mathop{\mathrm{SU}}\nolimits}
\newcommand{\SO}{\mathop{\mathrm{SO}}\nolimits}
\newcommand{\Sp}{\mathop{\mathrm{Sp}}\nolimits}
\newcommand{\UU}{\mathop{\mathrm{U}}\nolimits}

\newtheorem{remark}{Remark}

\newcommand{\microsection}[1]{\noindent\underline{\textbf{#1}}}
\newcommand{\Ref}[1]{(\ref{#1})}
\newcommand{\blanc}{\ifthenelse{\boolean{@twoside}}{\thispagestyle{empty}~\newpage}{} }

\begin{document}


\thispagestyle{empty}

\makeatletter
\begin{titlepage}
\begin{flushright}
{hep-th/0511132}\\
\end{flushright}
\vskip 1cm
\begin{center}
{\LARGE\bf Cubic curves from instanton counting}
\vskip 2cm
{\bf Sergey Shadchin}
\end{center}
\vskip 1cm
\centerline{\em INFN, Sezione di Padova \& Dipartimento di Fisica ``G. Galilei''}
\centerline{\em Universit\`a degli Studi di Padova, via F. Marzolo 8, Padova, 35131, ITALY}
\centerline{\tt email: serezha@pd.infn.it}
\vskip 3cm
\centerline{\sc Abstract}
\vskip 1cm
We investigate the possibility to extract Seiberg-Witten curves from the formal series for the prepotential, which was obtained by the Nekrasov approach. A method for models whose Seiberg-Witten curves are not hyperelliptic is proposed. It is applied to the $\SU(N)$ model with one symmetric or antisymmetric representations as well as for $\SU(N_1)\times \SU(N_2)$ model with $(N_1,N_2)$ or $(N_1,\cc{N}_2)$ bifundamental matter. Solution are compared with known results. For the gauge group product we have checked the instanton corrections which follow from our curves against direct instanton counting computations up to two instantons. 
\bigskip
\end{titlepage}
\makeatother


\tableofcontents


\section{Introduction}

String theory can shed some light to the strong coupling regime in the supersymmetric gauge theories. In particular the $\N=2$ super Yang-Mills theory is believed to be described by the type IIA superstrings NS5-D4 branes setup \cite{PrepFromM}. The Coulomb branch of this theory in the low-energy sector is described by a complex function on the moduli space of the theory, known as \emph{prepotential} \cite{Prepotential}. A very elegant construction for this function was proposed by Seiberg and Witten \cite{SeibergWitten,SeibergWittenII}. This construction includes a Riemann surface (the Seiberg-Witten curve) and a differential $\l(x)$ defined on this surface. The prepotential $\F(a)$ can be defined indirectly with the help of relations \Ref{A-cycles}. The crucial observation made in \cite{PrepFromM} is that the NS5-D4 type IIA setup can be lifted to M-theory there it becomes a single object, M5-brane, wrapped around a two-dimensional space, which can be associated with the Seiberg-Witten curve. 

This point of view yields to the solutions for numerous models, such as models with gauge group product \cite{PrepFromM,GroupProductFromBranes} symplectic and orthogonal group \cite{N=2BranesOrientifolds,NewCurvesFromBranes,MtheoryAndSW-SOandSp}, and symmetric and antisymmetric representation for unitary group \cite{NewCurvesFromBranes}. 

Three years ago another way to solve $\N=2$ super Yang-Mills theory was proposed by Nekrasov \cite{SWfromInst}. It is based on the localization technique, which, together with a certain deformation of the theory, gives the direct access to the prepotential after the explicit summation over the instanton contributions. This method (\emph{instanton counting}) yields the prepotential already as a series on the dynamically generated scale, without hard cycle integration of the Seiberg-Witten theory. However to study such effects as confinement or monopole condensation we have to know how to continue the prepotential beyond the convergence radius of proposed series. Seiberg-Witten theory can solve this problem. Therefore we have faced with the question how to extract the Seiberg-Witten geometry from the series for the prepotential. Also having found the Seiberg-Witten curve we gain an independent test of solutions, obtained by other methods. In particular we get a test for the M-theory.

In \cite{SWandRP} this problem was solved by the conformal map method, and the curves extracting technology was generalized in \cite{ABCD,SPinSW,MyThesis} to other groups and matter content. It was shown that the instanton counting defines some singular equations (saddlepoint equations) which enable us to find the Seiberg-Witten curve and differential. 

Conformal map method allows, however, to find the Seiberg-Witten curves (up to some rare exceptions) only in the case when curves are hyperelliptic. For more general situations it is not clear how to apply it. Thus pragmatically we need just another method to solve saddlepoint equations in order to get more Seiberg-Witten curves.

In this paper we propose such a method. It works well for the cubic curves, and, probably, can be generalized to other cases when the Seiberg-Witten curve is given by a finite degree polynomial. We consider the $\SU(N)$ model with symmetric or antisymmetric matter. Also, to elaborate more examples and check the curve predictions with the instanton counting predictions we have describe the instanton counting for the gauge group product.

The paper is organized as follows. In Section \ref{GGproduct} we describe the generalization of the Nekrasov approach for the gauge group product. Also we compute one- and two-instanton corrections to the prepotential for $\SU(N_1)\times \SU(N_2)$ model with one bifundamental matter representation of type $(N_1,\cc{N}_2)$ or $(N_1,N_2)$. In Section \ref{ProductEq} we propose a method which can be used to solve saddlepoint equations. As an illustration we apply it to the hyperelliptic curve models. In Section \ref{Cubic} we solve these equation for the symmetric and antisymmetric representations of $\SU(N)$ as well as for the $\SU(N_1)\times \SU(N_2)$ case with two types of  bifundamental matter. In Section \ref{Discussion} we discuss obtained results and check curve predictions against the instanton counting predictions.


\microsection{Acknowledgments.} I am very grateful to Nikita Nekrasov for numerous fruitful discussions. I would like to thank Ivan Kostov for his patient explanation of the singular equations structure. Also I thank all my colleagues of Dipartimento di Fisica to give me an opportunity to present my work as soon as it becames presentable. This work was partially supported by the EU MRTN-CT-2004-005104 grant ``Forces Universe'' and by by the MIUR contract no. 2003023852.


\section{Instanton counting for group product}
\label{GGproduct}

In this section we generalize the Nekrasov approach for the case when the gauge group is a direct product of simple classical groups. We will mostly use notations introduced in the Appendix \ref{InstCount}, which contains a brief survey of the instanton counting methods. It is not our ambition to exhaust all possible cases. For illustrative purposes it is sufficient to consider the product of two unitary groups. Generalization to other classical groups with richer matter content is straightforward \cite{SPinSW,MyThesis}.


\subsection{$\SU(N_1)\times \SU(N_2)$ case}

We consider the simplest case $G = \SU(N_1) \times \SU(N_2)$. When we deal with the product of two groups the general expression for the partition function, which generalizes \Ref{Partition} and \Ref{PartitionInst}, depends on two dynamically generated scales $\gL_1$ and $\gL_2$:
$$
\< 1 \>_a = Z^{pert}(a,m,\gL_1,\gL_2;\eps) Z^{inst}(a,m,\gL_1,\gL_2;\eps) = \exp\frac{1}{\eps_1\eps_2} \F(a,m,\gL_1,\gL_2;\eps), 
$$
where
\begin{equation}
\label{PartInst2}
Z^{inst}(a,m,\gL_1,\gL_2;\eps) = \sum_{k_1=0}^\infty \sum_{k_2=0}^\infty q_1^{k_1} q_2^{k_2}\oint \prod_{i=1}^{k_1} \frac{\dd \phi_i}{2\pi i}\oint \prod_{j=1}^{k_2} \frac{\dd \vf_j}{2\pi i} z_{k_1,k_2}\left(a,m,\phi,\vf;\eps\right).
\end{equation}
Here $q_1 = \e^{ 2\pi i \t_1(\gL_1)} = \gL^{\b_1} \e^{ 2\pi i \t_0^{(1)}}$ and the same for $q_2$. $\b_1$ and $\b_2$ can be computed using the matter content of the theory. Now the prepotential and the partition function depend on two sets of Higgs vacuum expectations: $a_1,\dots,a_{N_1}$ and $b_1,\dots,b_{N_2}$. Also the integration is performed over two dual group maximal torus Lie algebras ($\phi$ and $\vf$). It reflects the fact that in the case of gauge group product the total dual group is also the product of corresponding dual groups.

If there is no matter multiplets in a  ``mixed'' representation (which has a non-trivial charge which respect to both groups) then the instanton partition function \Ref{PartInst2} factorizes and there are no new effects. 

We consider the simplest non-trivial example of such a ``mixed'' representation, the bifundamental one. There are two types of bifundamentals: $ (N_1, N_2)$, which will be referred in that follows as ``$+$'' and $(N_1,\cc{N}_2)$ which will be referred as ``$-$''. Let us study both of them. 

First we are going to find the equivariant index for the Dirac operator. It can be done at the same way as \Ref{IndexAdjoint}. We have (using \Ref{ChernFund} and \Ref{IndexFund})
$$
\begin{aligned}
\Ind_q^\pm &= \frac{1}{(\e^{i\eps_1} -1)(\e^{i\eps_2}-1)} \sum_{l=1}^{N_1}\sum_{p=1}^{N_2} \e^{ia_l \pm i b_p} -\sum_{i=1}^{k_1} \sum_{p=1}^{N_2} \e^{i\phi_i\pm ib_p-i\eps_+} \\
&-\sum_{j=1}^{k_2} \sum_{l=1}^{N_1} \e^{\pm i\vf_j+ia_l- i\eps_+} +(\e^{-i\eps_1} -1)(\e^{-i\eps_2} -1) \sum_{i=1}^{N_1}\sum_{j=1}^{N_2} \e^{i\phi_i\pm i\vf_j}.
\end{aligned}
$$

The integrand of the partition function is (recall that we have shifted the masses by $-\eps_+$, see \Ref{zFund})
\begin{equation}
\label{zBifund}
z_{k_1,k_2}^\pm (a,b,\phi,\vf,M;\eps) = \frac{1}{k_1!k_2!} \frac{D_\pm(M-\eps_-)D_\pm(M+\eps_-)}{D_\pm(M-\eps_+)D_\pm(M+\eps_+)} \prod_{i=1}^{k_1}\P_2(\mp\phi_i\mp M) \prod_{j=1}^{k_2} \P_1(\mp \vf_j-M)
\end{equation}
where 
$$
\begin{aligned}
D_\pm(x) &= \prod_{i=1}^{k_1}\prod_{j=1}^{k_2} \left(\phi_i\pm\vf_j + x \right), & \P_1(x) &= \prod_{l=1}^{N_1}(x-a_l), & \P_2(x) &= \prod_{p=1}^{N_2}(x-b_p).
\end{aligned}
$$
When $k_1=0$ or $k_2=0$ we have $D_\pm(x) = 1$. Note that the integrand is invariant under the following transformation
\begin{equation}
\label{symmetry}
\begin{aligned}
1&\leftrightarrow 2, & M &\leftrightarrow \pm M.
\end{aligned}
\end{equation}
It means that $z_{k_1,k_2}^\pm (a,b,\phi,\vf,M;\eps) = z_{k_2,k_1}^\pm (b,a,\vf,\phi,\pm M;\eps)$. 


\subsection{Instanton corrections}

Using the instanton counting strategy we can compute some instanton corrections for the group product. With the two-instanton accuracy we have for both types of the bifundamental the following expression for the partition function \Ref{PartInst2}:
$$
\begin{aligned}
Z^{inst}_\pm  &= 1 + q_1 Z_{1,0}^\pm + q_2 Z_{0,1}^\pm  + q_1^2 Z_{2,0}^\pm + q_2^2 Z_{0,2}^\pm + q_1 q_2 Z_{1,1}^\pm + \dots,
\end{aligned}
$$
where 
\begin{equation}
\label{Zk1k2}
Z_{k_1,k_2}^\pm = \oint \prod_{i=1}^{k_1} \frac{\dd \phi_i}{2\pi i}\oint \prod_{j=1}^{k_2} \frac{\dd \vf_j}{2\pi i} z_{k_1,k_2}^\pm(\phi,\vf)
\end{equation}
(in this section we do not display such arguments of $z_{k_1,k_2}(\phi,\vf)$  as $a$, $b$, $M$, etc).

Consider the model with one bifundamental matter multiplet $(N_1, N_2)$ (``$+$'') or $(N_1,\cc{N}_2)$ (``$-$'') and also $N_f^{(1)}$ fundamental matter of $\SU(N_1)$ with masses $m_f^{(1)}$, $f=1,\dots, N_f^{(1)}$, and $N_f^{(2)}$ fundamental matter of $\SU(N_2)$ with masses $m_f^{(2)}$. In this model we have $\b_1 = 2N_1 - N_2 - N_f^{(1)}$ and $\b_2 = 2N_2 - N_1 - N_f^{(2)}$. Using formulae for $z_k(\phi)$ for the fundamental and adjoint representations \cite{SPinSW,MyThesis} (which can be obtained from \Ref{IndexFund} and \Ref{IndexAdjoint} with the help of \Ref{zFromInd}) as well as \Ref{zBifund} we gain functions $z_{k_1,k_2}(\phi,\vf)$. Plugging them into \Ref{Zk1k2} we can perform the integration and obtain the instanton corrections. With two-instanton accuracy the result is (in fact, $Z_{1,0}$ and $Z_{2,0}$ are the same for one unitary group model with specific fundamental matter content. Therefore we can take the expression from \cite{SWfromInst})
$$
\begin{aligned}
\hbar^2 Z^\pm_{1,0} &= -\sum_{l=1}^{N_1} S_l^\pm(0),\\
\hbar^4 Z^\pm_{1,1} &= \sum_{l=1}^{N_1}\sum_{p=1}^{N_2}S_l^\pm(0) T_p^\pm(0)\left( 1 - \frac{\hbar^2}{(a_l \pm b_p+M)^2} \right) \\
\hbar^4 Z_{2,0}^\pm &= \frac{1}{2}\sum_{l\neq m}^N \frac{S_l^\pm(0)S_m(0)^\pm}{\left(1- \frac{\hbar^2}{(a_l-a_m)^2} \right)^2} +\frac{1}{4}\sum_{l=1}^N S_l^\pm(0)\Big(S_l^\pm(\hbar) + S_l^\pm(-\hbar) \Big).
\end{aligned}
$$
where
\begin{equation}
\label{STdef}
\begin{aligned}
S^\pm(x) &= \frac{Q_1(x)\P_2(\mp x \mp M)}{\P_1^2(x)}, & S_l^\pm(x) & = \frac{Q_1(a_l + x)\P_2(\mp a_l \mp x \mp M))}{\prod_{l\neq m}(a_l-a_m +x)^2}, \\
T^\pm(x) &= \frac{Q_2(x)\P_1(\mp x - M)}{\P_2^2(x)}, & T_p^\pm(x) & = \frac{Q_2(b_p + x)\P_1(\mp b_p \mp x - M))}{\prod_{p\neq q}(b_p-b_q +x)^2}, \\
Q_1(x)&= \prod_{f=1}^{N_f^{(1)}} \left(x + m_f^{(1)}\right), & Q_2(x) &= \prod_{f=1}^{N_f^{(2)}} \left(x + m_f^{(2)}\right)
\end{aligned}
\end{equation}
$S^\pm(x)$ and $T^\pm(x)$ are referred as \emph{residue functions} \cite{EnnesMasterFunc,MTheoryTested}, $S^\pm_l(x)$ and $T^\pm_p(x)$ being their ``residues''. $T_p^\pm(x)$ can be obtained from $S_l^\pm(x)$ after the transformation \Ref{symmetry}. Note that to compute $Z_{1,1}^\pm $ we have used the fact that for  the particular choice \Ref{STdef} the following identities hold $T^\pm(\mp a_l \mp M) = S^\pm(\mp b_p - M) = 0$.

Corresponding series for the instanton part of the prepotential is
$$
\F^{inst}_\pm = q_1\F^\pm_{1,0}  + q_2 \F^\pm_{0,1} +q_1^2 \F^\pm_{0,2} + q_2^2\F^\pm_{2,0} + q_1q_2 \F^\pm_{1,1} + \dots,
$$
where
\begin{equation}
\label{PrepSUxSU}
\begin{aligned}
\F^\pm_{1,0} &= \sum_{l=1}^{N_1} S_l^\pm(0), \\
\F^\pm_{2,0} &= - \sum_{l\neq m}^{N_1} \frac{S_l^\pm(0)S_m^\pm(0)}{(a_l-a_m)^2} - \frac{1}{4} \sum_{l=1}^{N_1} S_l^\pm(0){S_l^\pm}''(0), \\
\F^\pm_{1,1} &= \sum_{l=1}^{N_1}\sum_{p=1}^{N_2} \frac{S_l^\pm(0)T_p^\pm(0)}{(a_l \pm b_p +M)^2}.
\end{aligned}
\end{equation}
One can easily check that the Seiberg-Witten prepotential is also invariant under \Ref{symmetry}. Using this observation we can restore $\F^\pm_{0,1}$ and $\F^\pm_{0,2}$. 


\subsection{Thermodynamical limit for group product}
\label{ThermoProduct}

Let us describe in some details the passage to the thermodynamical limit $\eps_1$, $\eps_2\to 0$. We generalize the results announced in Section \ref{Thermo}.

The double sum \Ref{PartInst2} is dominated by a single term with $\ds k_1 \sim k_2 \sim \frac{1}{\eps_1\eps_2}$. Since now we have two dual groups it is natural to introduce two profile functions:
$$
\begin{aligned}
f_1(x) &= \sum_{l=1}^{N_1} |x-a_l| -2\eps_1\eps_2 \sum_{i=1}^{k_1} \d(x-\phi_i), & f_2(x) &= \sum_{p=1}^{N_2} |x-b_p| -2\eps_1\eps_2 \sum_{j=1}^{k_2} \d(x-\vf_j).
\end{aligned}
$$
The Hamiltonian for the bifundamentals ``$+$'' and ``$-$'' (the ``interaction term'') is given by (we have taken into account that the matter is described by fermionic functions, therefore the sign is changed)
\begin{equation}
\label{Bifund}
H^\pm[f_1,f_2] = \frac{1}{4}\int \dd x\dd y f_1''(x)\k(x\pm y + M) f_2''(y).
\end{equation}

As in the single group case, the dependence on $\gL_1$ and $\gL_2$ is introduced via the following term in the total Hamiltonian of the model:
$$
+\frac{\pi i}{2} \left(\t_1(\gL_1) \int \dd x f_1''(x) x^2 + \t_2(\gL_2) \int \dd x f_2''(x) x^2 \right).
$$
In our situation when minimizing the free energy of the theory we obtain a couple of equations instead of single one \Ref{SP}:
\begin{equation}
\label{SP2}
\left\{
\begin{aligned}
\frac{1}{\pi i}\frac{\d H[f_1,f_2]}{\d f_1'(t)} &= \xi_l + t \t_1(\gL_1), & t&\in \g_l, & l &= 1,\dots,N_1, \\
\frac{1}{\pi i}\frac{\d H[f_1,f_2]}{\d f_2'(t)} &= \eta_p + t \t_2(\gL_2), & t&\in \d_p, & p &= 1,\dots,N_2,\\
\end{aligned}
\right.
\end{equation}
where $\g_l$ and $\d_p$ are cuts for two groups, and $\xi_l$, $\eta_p$ are certain constants, in general different for different cuts. $\ds \t_1(\gL_1) = \t_0^{(1)} + \frac{1}{2\pi i} \ln \gL_1^{\b_1}$ and the same for $1\leftrightarrow 2$. In order to solve these equations we introduce the primitives of the profile function resolvents as follows
$$
F_{1,2}(z) = \frac{1}{4\pi i} \int \dd x f_{1,2}''(x) \ln (z-x).
$$
Then the equations \Ref{SP2} can be rewritten as difference equations for these functions. The prepotential is defined indirectly by formulae which generalize \Ref{A-cycles}:
$$
\begin{aligned}
\oint_{A_l} z F_1'(z) \dd z &= a_l, & 2 \pi i \oint_{B_l}  z F_1'(z) \dd z= \frac{\pd \F}{\pd a_l} = a^l_D \\
\oint_{C_p} z F_2'(z) \dd z &= b_p, & 2 \pi i\oint_{D_p}  z F_2'(z) \dd z= \frac{\pd \F}{\pd b_p} =  b^p_D.
\end{aligned}
$$
The unusual property is that now we have a \emph{couple} of Seiberg-Witten differentials: $\l_{1,2}(z) = z F_{1,2}'(z)\dd z$. It seems to be in the opposition with Seiberg-Witten consideration, but as we shall see soon, in our examples $\l_1(z)$ and $\l_2(z)$ are not independent, and therefore we have only one differential.


\section{Product equations}
\label{ProductEq}

The difference equations for $F_{1,2}(z)$ which follow from  \Ref{SP2} define the Seiberg-Witten curve as well as the Seiberg-Witten differential \cite{SWandRP,SPinSW,MyThesis}. The conformal map method, proposed in \cite{SWandRP}, is powerful enough when we deal with hyperelliptic curves. It corresponds to the Yang-Mills theories with some fundamental matter multiplets. Solutions of more general difference equations can not be obtained likewise (except for some very particular exceptions). In this section we propose another method which allows to find solutions for more general models. Namely having exponentiated a difference equation we obtain a \emph{product equation}. It turns out that the Vieta theorem together with the simplicity principle allows to determine completely particular solutions of product equations. We did not try to prove the unicity of obtained solutions. Instead we have found them for various models and checked for consistency.


\subsection{An example}
\label{Example}

To exhibit the idea let us consider in some details the simplest case: the Yang-Mills theory for the group $\SU(N)$ with $N_f < 2N$ fundamental matter multiplets. The difference equation constructed with the help of the Table \ref{Diff} is given by the expression \Ref{DifferenceExample}. The running complex coupling constant in this example is $\ds \t(\gL) = \t_0 + \frac{1}{2\pi i} \ln \gL^{2N-N_f}$. Since our theory is not conformal  we can neglect the first term and put simply $\ds \t(\gL) = \frac{1}{2\pi i} \ln \gL^{2N-N_f}$. In that follows we will basically drop $\t_0$.

In the Seiberg-Witten theory one works not with $F(z)$, but rather with its exponent $y(z)$ defined in \Ref{yDef}. Let us rewrite the difference equation in terms of this function. Taking the exponent we obtain the \emph{product equation}:
\begin{equation}
\label{product}
\begin{aligned}
y^+(t) y^-(t) &= q Q(t), & t &\in \g_l, 
\end{aligned}
\end{equation}
where $q = 2\pi i \t = \gL^{2N-N_f}$ is the instanton counting parameter and $\ds Q(z) = \prod_{f=1}^{N_f}(z + m_f)$.

Note that the properties of the profile function \Ref{profileProp} imply that
\begin{equation}
\label{Fassym}
F(z) = \frac{N}{2\pi i} \ln z + O\left(\frac{1}{z^2}\right)
\end{equation}
when $z\to\infty$. Therefore in this limit we have $y(z) = z^N + O(z^{N-2})$. Such a  behavior can take place if $y(z)$ is a solution of an algebraic equation with $z$-dependent coefficients. Suppose this is the case. Let the degree of this polynomial be $n$. It follows, that $y(z)$ is one of its roots, which we denote as $y_1(z),\dots,y_n(z)$. This equation defines an algebraic curve. Suppose as well that this curve possess only double ramification points. That is, at a particular value of $z$ no more than two roots can coincide. This statement is justified in Appendix \ref{doubleRamification}.

Let us come back to \Ref{product}. Without loss of generality we can suppose that $y_1^+(t)$ is what we mean by $y^+(t)$ and $y_2^+(t) = y_1^-(t)$ is $y^-(t)$  when $t\in\g_l$. Therefore for our model we find that $y_1^+(t)y_2^+(t) = qQ(t)$ on the cuts.  Since $y_1(z)$ and $y_2(z)$ are holomorphic functions, we can continue the product equation from cuts to the whole domain of their analyticity. Therefore these functions satisfy $y_1(z)y_2(z) = qQ(z)$. The simplest equation with at least two roots is the quadratic equation. With the help of the  Vieta theorem we conclude that the desired equation for $y(z)$ looks like
$$
y^2(z) - P(z) y(z) + qQ(z) = 0,
$$
where $P(z)$ is a polynomial of $z$. Further analysis shows that the cuts appears around the zeros of the polynomial $P(z)$, and therefore for the $\SU(N)$ model we should use a degree $N$ polynomial. Conventionally it is written as $\ds P(z) = \prod_{l=1}^N(z-\a_l)$ with some parameters $\a_l$ which are related to the Higgs expectation values via \Ref{A-cycles}. The condition \Ref{Fassym} shows that for $y(z)$ which defines $F(z)$ we should take the following root (the branch of the square root is defined in such a way that $\sqrt{1 + 2z} \approx 1 + z$):
$$
y(z) = \frac{P(z)}{2} \left(1 + \sqrt{1 - \frac{4qQ(z)}{P^2(z)}} \right).
$$
 

\subsection{Symplectic case}
\label{SymplecticCase}

In that follows it will be useful to discuss some aspects of $\Sp(N)$ models with $N_f$ fundamental matter.  The difference equation for the primitive of the profile function resolvent can be deduces form the Table \ref{Diff}. We get
$$
\begin{aligned}
F^+(t) + F^-(t) + \frac{2}{\pi i} \ln |t| - \frac{1}{4\pi i} \sum_{f=1}^{N_f} \Big(\ln|t+m_f| + \ln|-t+m_f| \Big)&= 2\t, & t&\in \g_l,
\end{aligned}
$$
where $2\pi i \t = \ln \gL^{N + 1 - N_f}$. As it was shown in \cite{ABCD} the profile function for the this case is symmetric. It follows that $y(z) = y(-z)$. In order to absorb the second term we redefine the profile function as follows: $\tld{f}(x) = f(x) + 2|x|$. According to the profile function redefinition we also introduce
$$
\begin{aligned}
\tld{F}(z) &= \frac{1}{4\pi i}\int \dd x \tld{f}''(x)\ln (z-x) = F(z) + \frac{1}{\pi i} \ln z, \\
\tld{y}(z) &= \exp2\pi i \tld{F}(z) = y(z) z^2. 
\end{aligned}
$$

Then for the new function we get
$$
\begin{aligned}
\tld{F}^+(t) + \tld{F}^-(t) - \frac{1}{2\pi i}\sum_{f=1}^{N_f} \Big(\ln |t + m_f|+\ln |t - m_f| \Big) &= 2\t, & t&\in \pm \g_l.
\end{aligned}
$$

Formally this equation looks like the equation for the  $\SU(2N+2)$ model with the following Higgs vevs: $\pm a_l$, $l=1,\dots,N$, $a_{2N+1} = a_{2N+2} = 0$. Moreover each fundamental matter multiplet with the mass $m$ is equivalent to a couple of $\SU(2N+2)$ fundamental multiplets with masses $\pm m$. 
The conformal map method together with the reflection principle allows us to find the Seiberg-Witten curve in this case \cite{ABCD,MyThesis}. The result is the following
\begin{equation}
\label{SpCurve}
\tld{y}^2(z) - \tld{P}(z) \tld{y}(z) + q^2 Q(z){(-1)}^{N_f}Q(-z) = 0,
\end{equation}
where 
\begin{equation}
\label{PforSp}
\tld{P}(z) = z^2 \prod_{l=1}^N(z^2-\a_l^2) + 2i^{N_f}q Q(0).
\end{equation}
Such a complicated form of this polynomial has, however, a very natural explanation. Indeed, in spite of the fact that we have taken as a starting point the curve for the $\SU(2N+2)$ theory, only $2N$ of this curve moduli are defined via \Ref{A-cycles}. Two others are known exactly (and are equal to zero). It means that there is no cut around the point $z = 0$. Otherwise, the cut that might appear in fact shrinks to the point. 

Mathematically this fact can be expressed as  statement that two roots of the quadratic equation matches at the point $z=0$. Therefore we should have
$$
\tld{y}^2(0) - \tld{P}(0)\tld{y}(0) + q^2{(-1)}^{N_f} Q^2(0) = {\Big( \tld{y}(0) - i^{N_f}qQ(0) \Big)}^2.
$$
Together with the symmetry of this polynomial under reflection $\tld{P}(z) = \tld{P}(-z)$ this condition determines completely $\tld{P}(z)$. It is straightforward to check that the discriminant of the quadratic equation \Ref{SpCurve} is proportional to $z^2$, that means that the cut around zero indeed shrinks to a single point.

Note also, that one might be worried about the appearance of a new singularity at the point $z = 0$. Even though the cut is shrinked to the point, the Seiberg-Witten differential is not holomorphic there, but rather has a pole. One checks, however, that when we get back to $y(z)$ from $\tld{y}(z)$ this pole disappears and the differential $\l(z) = z F'(z) \dd z$ has no singularities at $z = 0$. 


\section{Getting cubic curves}
\label{Cubic}

In this section we apply the logic developed previously to the models whose Seiberg-Witten curves are not hyperelliptic. As we shall see, the Vieta theorem supplemented by a symmetry principle allows to solve product equations for the cubic curves models. Strictly speaking, these curves should not be called ``cubic'', but rather ``trigonal''\footnote{I am grateful to A. Gorinov for this remark}, since they are defined by a polynomial which is cubic on $y$ and arbitrary degree on $z$ \cite{Curves}. However, following the tradition, we continue to call them so.


\subsection{Antisymmetric matter, special case}

Consider the theory with massive matter multiplet in the symmetric or antisymmetric representation with the mass $M$ and also $N_f$ fundamentals with masses $m_1,\dots ,m_{N_f}$. With the help of the Table \ref{Diff} we get the following saddlepoint equation for the profile function:
\begin{multline}
\label{Ant}
\begin{aligned}
F^+(t) + F^-(t) - \frac{1}{4\pi i}\sum_{f=1}^{N_f} \ln |t + m_f| \pm \frac{1}{\pi i} \ln \left| t + \frac{M}{2} \right|&=   F(-t-M) + \t, & t &\in \g_l.
\end{aligned}
\end{multline}
where ``$+$'' is taken for antisymmetric representation and ``$-$'' for the symmetric one, $2\pi i\t = \ln \gL^{N\pm 2-N_f}$.

Now let us consider the special case: antisymmetric matter with the mass $M$ and two fundamental multiplets with masses $M/2$. In this special case $\b = N$. The difference equation for $F(z)$ simplifies and we obtain
$$
\begin{aligned}
F^+(t) + F^-(t) &= F(-t-M) + \t, & t &\in \g_l.
\end{aligned}
$$
The product equation for $y(z)$ is, therefore,
$$
\begin{aligned}
y^+(t)y^-(t) &= qy(-M-t), & & t\in \g_l.
\end{aligned}
$$
Let us try to find a solution for this equation in spirit of the previous discussion. Associate, as before, $y^+(t)$ with $y_1^+(t)$ and $y^-(t)$ with $y_1^-(t)=y_2^+(t)$. Moreover suppose that there is another root of the algebraic equation, $y_3(z)$, which satisfies 
\begin{equation}
\label{RootCondition}
y_3(-M-z) = \frac{q^2}{y_1(z)}
\end{equation}
If it is so, we obtain (after an analytical continuation to the whole domain of analyticity)
$$
y_1(z)y_2(z)y_3(z) = q^3.
$$

The simplest equation which has at least three roots is the cubic one. Let us try it. By the Vieta theorem we get
$$
y^3(z) - P(z) y^2(z) - R(z) y(z) + q^3 = 0.
$$
The condition \Ref{RootCondition} implies that the curve should have the following ``reflection symmetry'': 
$$
y(-z-M) = \frac{q^2}{y(z)}.
$$
However the converse is not true: if the curve has such a symmetry, \Ref{RootCondition} might not be satisfied. We can only state that there is \emph{a} root $y_a(z)$ such that $\ds \frac{q^2}{y_1(z)} = y_a(-M-z)$. If it were $y_2(t)$, for example, the method did not work. Hopefully, the analysis of cubic equations roots \cite{EnnesSUSymFund} shows that we are away of troubles.

It follows that $R(z) = qP(-z-M)$. The analysis of cuts shows that for $\SU(N)$ theory ($N$ cuts) the degree of $P(z)$ is $N$. Therefore we get the following curve:
$$
y^3(z) - P(z) y^2(z) - q P(-z-M) y(z) +q^3 = 0
$$
which solves the product equation.


\subsection{Symmetric matter}

Prior to discuss the the symmetric matter let us first consider an extension of the previous case, the antisymmetric matter with $N_f$ fundamentals with masses $m_1 = M/2$, $m_2=M/2$ and other masses being arbitrary. The product equation is
\begin{equation}
\label{FacSym}
\begin{aligned}
y^+(t)y^-(t) &= q y(-t-M)\tld{Q}(t) & t&\in \g_l,
\end{aligned} 
\end{equation}
where $\ds \tld{Q}(z) = \prod_{f=3}^{N_f} (z+m_f)$ and $q = \gL^{N+2-N_f}$. 

In order to find the solution we suppose that the equation is cubic and that it possess a symmetry which generalizes \Ref{RootCondition}
\begin{equation}
\label{RootCondGen}
\ds y_1(-z-M) = \frac{A(z)}{y_3(z)}.
\end{equation}
Then we get
$$
y_1(z)y_2(z)y_3(z) = q\tld{Q}(z) A(z).
$$
The Vieta theorem implies 
$$
y^3(z) - P(z) y^2(z) - R(z) y(z) + q\tld{Q}(z)A(z) = 0.
$$

Having applied the symmetry transformation we get:
\begin{multline*}
y^3(z) - \frac{R(-z-M) A(z)}{q\tld{Q}(-z-M)A(-z-M)} y^2(z) - \frac{P(-z-M)A^2(z)}{q\tld{Q}(-z-M) A(-z-M)}y(z) \\
+ \frac{A^3(z)}{q\tld{Q}(-z-M)A(-z-M)} = 0.
\end{multline*}
Therefore we get the following conditions to determine $A(z)$ and $R(z)$:
\begin{equation}
\label{CoeffCond}
\begin{aligned}
&(i) &  P(z) &= \frac{R(-z-M) A(z)}{q\tld{Q}(-z-M)A(-z-M)}, \\
&(ii) & R(z) &= \frac{P(-z-M)A^2(z)}{q\tld{Q}(-z-M) A(-z-M)}, \\
&(iii) &\tld{Q}(z)A(z)q &= \frac{A^3(z)}{q\tld{Q}(-z-M)A(-z-M)}.
\end{aligned}
\end{equation}
The consistency of $(i)$ and $(ii)$ implies $\ds A(z) = q^2\tld{Q}(z)\tld{Q}(-z-M)$. Having put this result into $(ii)$ we obtain $R(z) = q P(-z-M)\tld{Q}(z)$. Note that at this stage we have no more freedom. However, the condition $(iii)$ is still untouched. We are lucky and having put $A(z)$ into $(iii)$ we see that it is automatically satisfied. Therefore the desired curve is
$$
y^3(z) - P(z)y^2(z) - qP(-z-M)\tld{Q}(z) y(z) + q^3\tld{Q}^2(z)\tld{Q}(-z-M) = 0.
$$

Finally to pass to the symmetric matter with $N_f$ fundamentals multiplet we note that the symmetric matter with $N_f$ fundamentals is equivalent to the antisymmetric matter with $N_f+4$ fundamentals with masses $m_1 = \dots = m_4 = M/2$ and the other masses being arbitrary \cite{SPinSW,MyThesis}. Having introduced $\ds Q(z) = \prod_{f=1}^{N_f}(z+m_f)$ we get the following curve:
\begin{equation}
\label{CurveSym}
y^3(z) - P(z) y^2(z) - qP(-z-M)Q(z){\left(z+\frac{M}{2}\right)}^2 y(z) + q^3Q^2(z)Q(-z-M){\left(z+\frac{M}{2}\right)}^6= 0,
\end{equation}
where now $q=\gL^{ N-2-N_f}$, as it should be for the symmetric matter.


\subsection{Antisymmetric matter}

Let us finally deduce the curve for the antisymmetric matter and $N_f$ fundamentals. Our strategy will be basically the same as in Section \ref{SymplecticCase}.

First we redefine the profile function in \Ref{Ant} as follows:
$$
\tld{f}(x) = f(x) + 2 \left|x+ \frac{M}{2}\right|.
$$
The difference equation for 
$$
\ds \tld{F}(z) = \frac{1}{4\pi i} \int \dd x \tld{f}''(x) \ln (z-x) = F(z) + \frac{1}{\pi i} \ln \left(z+\frac{M}{2}\right)$$ 
is
$$
\begin{aligned}
\tld{F}^+(t) + \tld{F}^-(t) &= \tld{F}(-t-M) +\t + \ln Q(t), & t&\in \g_l.
\end{aligned}
$$
Therefore for $\ds\tld{y}(z) = \exp 2\pi i \tld{F}(z) = y(z) {\left( z+ \frac{M}{2}\right)}^2$ we get the following product equation
$$
\begin{aligned}
\tld{y}^+(t)\tld{y}^-(t) &= q\tld{y}(-t-M) Q(t) , & t&\in\g_l.
\end{aligned}
$$

Formally it is the same as \Ref{FacSym}. Thus we can immediately write corresponding Seiberg-Witten curve:
\begin{equation}
\label{AntCurve}
\tld{y}^3(z) - \tld{P}(z) \tld{y}^2(z) -q \tld{P}(-z-M)Q(z) \tld{y}(z) + q^3Q^2(z)Q(-z-M) = 0.
\end{equation}

The subtle point is to determine the polynomial $\tld{P}(z)$. The general form of this polynomial is $\ds \tld{P}(z) = (z - \mu_1 )(z - \mu_2)\prod_{l=1}^N \left(z-\tld{\a}_l\right)$ where $\mu_1$ and $\mu_2$ are roots which in the perturbative approximation go to $-\frac{M}{2}$. As in symplectic case these parameters do not belong to the curve moduli space, but rather should be defined otherwise. Our guide principle will be, as in the symplectic case, the absence of the cut around $\ds z = - \frac{M}{2}$. It follows that at this point the cubic polynomial has all roots matched. It means that
\begin{multline*}
\tld{y}^3(-M/2) - \tld{P}(-M/2)\tld{y}^2(-M/2) - q\tld{P}(-M/2)Q(-M/2)\tld{y}(-M/2)+q^3 Q^3(-M/2) \\
= {\Big(\tld{y}(-M/2) + q Q(-M/2) \Big)}^3.
\end{multline*} 
Therefore $\tld{P}(-M/2) = - 3 q Q(-M/2)$. Note that it does not contradict to the statement announced in Section \ref{Example} that the endpoints of cuts are always double ramification points, since the statement concerns only cuts, defined by the saddlepoint equation.

Looking at \Ref{PforSp} it is natural to propose 
$$
\tld{P}(z) = {\left(z+\frac{M}{2}\right)}^2 \prod_{l=1}^N(z-\a_l) - 3q Q\left(-\frac{M}{2}\right) =  {\left(z+\frac{M}{2}\right)}^2 \prod_{l=1}^N(z-\a_l) - 3q \prod_{f=1}^{N_f}\left(m_f-\frac{M}{2}\right) .
$$
It is straightforward to check that the discriminant of the cubic equation \Ref{AntCurve} with this $\tld{P}(z)$ is proportional to $\ds \left(z+\frac{M}{2}\right)^2$. It means that the cut around $\ds z = - \frac{M}{2}$ shrinks to the point, as we demanded.

Exactly as in the symplectic case the pole that we have at $\ds z = -\frac{M}{2}$ for the Seiberg-Witten differential $\tld{\l}(z) = z \tld{F}'(z) \dd z$ disappears when we pass to $\l(z) = z F'(z) \dd z$.


\subsection{Curves for group product}

Cubic curves appear also in the theories with $\SU(N_1)\times \SU(N_2)$ gauge group \cite{PrepFromM}.  Let us see how does it work. 

Consider again the model with one bifundamental matter multiplet $(N_1, N_2)$ or $(N_1,\cc{N}_2)$ and some fundamental for both groups. The difference equations \Ref{SP2} which follow from the Hamiltonian \Ref{Bifund} for the bifundamntal of type ``$\pm$'' are (recall that $(N_1, N_2)$ is referred as ``$+$'' whereas $(N_1,\cc{N}_2)$ --- as ``$-$'')
$$
\left\{ 
\begin{aligned}
F_1^+(t) + F_1^-(t) - F_2(\mp t \mp M) - \frac{1}{2\pi i}\sum_{f=1}^{N_f^{(1)}} \ln|t+m_f^{(1)}| &= \t_1, & t&\in \g_l, \\
F_2^+(t) + F_2^-(t) - F_1(\mp t - M) - \frac{1}{2\pi i}\sum_{f=1}^{N_f^{(2)}} \ln|t+m_f^{(2)}| &= \t_2, & t&\in \d_p.
\end{aligned}
\right.
$$

At the same way as one introduces $y(z)$ in \Ref{yDef} we define a couple of functions 
$$
\begin{aligned}
y(z) &= \exp 2\pi i F_1(z), & w(z) &= \exp 2\pi i F_2(z).
\end{aligned}
$$
The product equations for them can be written as follows:
\begin{equation}
\label{FacG1G2}
\left\{
\begin{aligned}
y^+(t)y^-(t) &= q_1Q_1(t)w(\mp t\mp M), & t&\in \g_l, \\
w^+(t)w^-(t) &= q_2Q_2(t)y(\mp t-  M), & t&\in \d_p.
\end{aligned}
\right.
\end{equation}

In order to solve these  equations we suppose that $y(z)$ and $w(z)$ are solutions of similar cubic equations. More precisely, we set as before $y^+(t) = y^+_1(t)$ and $y^-(t) = y^-_1(t) = y^+_2(t)$, and the same for $w(z)$, and we suppose that there is a following version of symmetries \Ref{RootCondition} and \Ref{RootCondGen}:
\begin{equation}
\label{G1G2Root}
\begin{aligned}
w_1(\mp z \mp M) &= \frac{A_1(z)}{y_3(z)}, & y_1(\mp z-M) &= \frac{A_2(z)}{w_3(z)}.
\end{aligned}
\end{equation}
It follows that the coefficients of equations which determine $y(z)$ and $w(z)$ 
\begin{equation}
\label{G1G2Curve}
\begin{aligned}
y^3(z) - P_1(z)y^2(z) - R_1(z)y(z) + q_1Q_1(z)A_1(z) &= 0, \\ 
w^3(z) - P_2(z)w^2(z) - R_2(z)w(z) + q_2Q_2(z)A_2(z) &= 0,
\end{aligned} 
\end{equation}
satisfy certain conditions (analogue of \Ref{CoeffCond}). Note that the equation for $w(z)$ can be obtained from the equation for $y(z)$ after the transformation \Ref{symmetry}. This system of conditions turns out to be overdefined, but hopefully we can find a solution:
$$
\begin{aligned}
A_1(z) &=  q_1Q_1(z)q_2Q_2(\mp z \mp M), & A_2(z) &= A_1(\mp -M) = q_1Q_1(\mp z-M)q_2Q_2(z), \\
R_1(z) &= q_1 Q_1(\mp z)P_2(\mp z \mp M), & R_2(z) &= q_2Q_2(z)P_1(\mp z-M).
\end{aligned}
$$

Therefore for $y(z)$ and $w(z)$ we get the following curve:
\begin{equation}
\label{CurveBifund}
\begin{aligned}
y^3(z) - P_1(z) y^2(z) - q_1 Q_1(z)P_2(\mp z \mp M) y(z) + q_1^2 Q_1^2(z)q_2Q_2(\mp z \mp M) &= 0 \\
w^3(z) - P_2(z) y^2(z) - q_2 Q_2(z)P_1(\mp z - M) y(z) + q_2^2 Q_2^2(z)q_1Q_1(\mp z - M) &= 0.
\end{aligned}
\end{equation}
Here $\ds P_1(z) = \prod_{l=1}^{N_1}(z - \a_l )$, $\ds P_2(z) = \prod_{p=1}^{N_2} (z - \r_p)$ are polynomials which define cuts. In the classical limit $\a_l \to a_l$ and $\r_p \to b_p$.


\section{Discussion}
\label{Discussion}


\subsection{M-theory curves}

In some cases the M-theory arguments can provide the exact form of the Seiberg-Witten curve for certain models. Namely in \cite{PrepFromM} Witten showed that a stack of D4-branes stretched between NS5-branes in type IIA string theory lifted to the M-theory describes non-perturbative effects in the $\N=2$ super Yang-Mills theory which lives on the infinite worldvolume of D4-branes. From the M-theoretical point of view the NS5-D4 setup can be seen as a single M5 brane embedded in $\Real^{1,9} \times \Sphere^1$ in a complicated way. Namely the M5-brane is supposed to be wrapped around a two-dimensional Riemann surface, which can be identified with the Seiberg-Witten curve.

This point of view is proved to be powerful and already in \cite{PrepFromM} the expressions for the Seiberg-Witten curve for group product of form $\SU(N_1)\times \SU(N_2) \times \dots \times \SU(N_n)$ interacting by means of massless bifundamental multiplets of form $(N_a, \cc{N}_{a+1})$, $a = 1,\dots,n$ were found. The solution obtained by Witten missed the manifest dynamically generated  scales dependence. It was restored in \cite{GroupProductFromBranes} for the special case when $Q_2(z) = \dots = Q_{n-1}(z) = 1$. Compiling these results one can guess the curve for the model without any restrictions on $Q_a(z)$ (except for, of cause, the asymptotic freedom). In our notations the curve can be written as follows (we have replaces $y(z)\mapsto -y(z)$ and $q_a\mapsto -q_a$):
\begin{multline*}
0= y^{n+1}(z) - P_1(z)y^n(z) - q_1 Q_1(z)P_2(z) y^{n-1}(z) + q_1^2Q_1^2(z)q_2 Q_2(z) P_3(z) y^{n-2}(z) + \dots \\
= y^{n+1}(z) + \sum_{a = 1}^n {(-1)}^{\frac{a(a+1)}{2}}y^{n+1-a}(z) P_a(z)\prod_{b=1}^{a-1} {\Big(q_b Q_b(z)\Big)}^{a-b} + {(-1)}^{\frac{(n+1)(n+2)}{2}} \prod_{a=1}^n{\Big(q_a Q_a(z)\Big)}^{n+1-a}.
\end{multline*}
For $n=2$ this curve matches with the curve \Ref{CurveBifund} for the type ``$-$'' bifundamental and $M=0$. Presumably the whole expression for arbitrary $n$ can be deduced from the instanton counting. In fact, it is rather straightforward to obtain the appropriate generalization of the system of equations \Ref{FacG1G2} (see \Ref{nProd}). However we can not find a solution.

By considering orientifold planes the result of \cite{PrepFromM,GroupProductFromBranes} was generalized rapidly to include to the product also the orthogonal and symplectic groups \cite{N=2BranesOrientifolds,MtheoryAndSW-SOandSp,NewCurvesFromBranes}. In particular in \cite{NewCurvesFromBranes} the curves for the single $\SU(N)$ with one symmetric or antisymmetric multiplet was proposed. They match with \Ref{CurveSym} and \Ref{AntCurve} respectively after the shift of all Higgs vevs by $\ds \frac{M}{2}$ and for the case $Q(z) = 1$ (no supplementary fundamental matter).

Unfortunately, there are no available (at least for author) results concerning the gauge group product interacting via type ``$+$'' bifundamental representation $(N_1, N_2)$. Therefore the curve \Ref{CurveBifund} with the sign ``$+$'' for the moment can not be tested M-theoretically. 


\subsection{Instanton corrections}
\label{InstantonCheck}

Another way to check our curves is to compare some low-instanton corrections obtained from them with direct instanton counting results. A considerable effort was devoted to extracting one- and two-instanton corrections from the Seiberg-Witten curves in the non-hyperelliptic case \cite{NaculichSUAntFund,EnnesSUSymFund,EnnesSUSymAnt,1instProduct,EnnesSU2AntFund,EnnesMasterFunc,MartaSUxSU}. In \cite{WyllMar} these results for the $\SU(N)$ model with one symmetric or antisymmetric matter multiplet were checked against the instanton counting formulae with two-instanton accuracy. Also in \cite{SPinSW,MyThesis} the hyperelliptic approximation for the Seiberg-Witten curves was obtained as an approximative solution of the saddlepoint equations. It was argued then that this approximation is sufficient to get the correct one-instanton results. Note also that in \cite{Quiver} the prepotential for the gauge group product with the bifundamental matter was obtained using the quiver gauge theory approach.

Now let us check the instanton correction for the gauge group product. We are going to compare explicit expressions \Ref{PrepSUxSU} for the  prepotential with instanton corrections  which can be obtained from Seiberg-Witten curves. Generally the computations are extremely difficult (especially for the $B$-cycles). Luckily there are some methods which allow to compute the prepotential without doing dual periods. For example method based on the Whitham hierarchy that arises from the identification of Seiberg-Witten solution and integrable models provides a recursive scheme, which does not require even $A$-cycles \cite{Gorsky,MarcoRecursion,JoseMarcos,JoseMarcosMarta}. Another method, which is based on the non-perturbative RG-equation fot the prepotential \cite{RGequation,RecursionEqn,RGMatone,PrepotentialEquation,prepotRelation} requires no identification with integrable models, but only the knowledge of Seiberg-Witten curve. We are going to use it, since this is the curve that we wish to test. 

In \cite{MartaSUxSU} the recursive relation method, proposed in \cite{RecursionEqn} was generalized to the product group case. The result is the following. Consider a general cubic curve which can be put into form \Ref{CurveBifund}. Then the prepotential can be expressed as follows (we have changed signs for two-instanton terms, which corresponds to $\F \mapsto - \F$, $q_{1,2} \mapsto - q_{1,2}$):
$$
\begin{aligned}
\F_{1,0} &= -\sum_{l=1}^{N_1} a_l \gD_l^{(1)}(0), \\
\F_{2,0} &= \frac{1}{2} \sum_{l=1}^{N_1} \left( a_l \gD_l^{(2)}(0) + \frac{1}{2}\gD_l^{(1)}(0)\gD_l^{(1)}(0) + \frac{\pd \F_{1,0}}{\pd a_l} \gD_l^{(1)}(0) \right), \\
\F_{1,1} &= \frac{\b_2}{\b_1+\b_2} \sum_{l=1}^{N_1} \frac{\pd \F_{0,1}}{\pd a_l} \gD_l^{(1)}(0) + \frac{\b_1}{\b_1+\b_2} \sum_{p=1}^{N_2} \frac{\pd \F_{1,0}}{\pd b_p} \gG_p^{(1)}(0),
\end{aligned}
$$
where
$$
\begin{aligned}
\gD_l^{(k)}(x) &= \frac{1}{(k!)^2} {\left(\frac{\pd}{\pd x} \right)}^{2k-1} S_l^k(x), & \gG_p^{(k)}(x) &= \frac{1}{(k!)^2} {\left(\frac{\pd}{\pd x} \right)}^{2k-1} T_p^k(x).
\end{aligned}
$$
For example $\gD_l^{(1)}(x) = S_l'(x)$, $\ds \gD_l^{(2)}(x) = \frac{3}{2} S_l'(x)S_l'(x) + \frac{1}{2}S_l(x)S_l''(x)$. The prepotentials $\F_{0,1}$ and $\F_{0,2}$ can be obtained after the transformation \Ref{symmetry}.

Let us prove that this is, in fact, the same as \Ref{PrepSUxSU}. First, using the identity \cite{HKP-SU}
$$
\res_{x=\infty} x S(x) = \sum_{x_0:\mr{finite}} \res_{x = x_0} x S(x) = \sum_{l=1}^{N_1} \left( a_l S_l'(0) + S_l(0)\right) = \sum_{l=1}^{N_1} \left( a_l \gD_l^{(1)}(0) + S_l(0)\right),
$$
where $\ds \res_{x=\infty} xS(x) = 0$, if $\b_1 > 1$, and taking into account the structure of $P_1(x)$ and $P_2(x)$ ($P(x) = x^N + u_2 x^{N-2}+\dots$) we see that if $\b_1 = 1$ then $\ds \res_{x=\infty} xS(x) = \sum_{f=1}^{N_f^{(1)}} m_f^{(1)} \pm M$, which is a non-physical constant which can be discarded. Nevertheless, for the conformal theories the residue at infinity becomes moduli dependent, and we can not neglect it. Since we do not consider conformal theories, we can use $\ds \sum_{l=1}^{N_1} a_l \gD_l^{(1)}(0) = - \sum_{l=1}^{N_1} S_l(0)$ and therefore we obtain the agreement for $\F_{1,0}$ and $\F_{0,1}$.

To test $\F_{2,0}$ we use the same trick. First we note that for the non-conformal theories ($\b > 0$) up to a nonphysical constant 
$$
0 = \res_{x = \infty} x S^2(x) = \sum_{l=1}^{N_1} \left( S_l'(0)S_l'(0) + S_l S_l''(0) + \frac{2}{3} a_l \gD_l^{(2)}(0) \right).
$$
Since
$$
\begin{aligned}
\frac{\pd S_l(x)}{\pd a_l} &= \frac{\pd S_l(x)}{\pd x} = S_l'(x), & \frac{\pd S_m(x)}{\pd a_l} &= \frac{2 S_m(x)}{a_m - a_l + x}, & \mbox{if } m&\neq l,
\end{aligned}
$$
we have
$$
\frac{\pd \F_{1,0}}{\pd a_l} = \sum_{m=1}^{N_1} \frac{\pd S_m(0)}{\pd a_l} = S_l'(0) + 2 \sum_{m\neq l}^{N_1} \frac{S_m(0)}{a_m - a_l}.
$$
Also we note that for non-conformal theories the following identity holds
\begin{equation}
\label{S/(x-a)}
0 = \res_{x = \infty} \frac{S(x)}{x-a} = S(a) - \sum_{m=1}^{N_1} \frac{S_m(0)}{(a_m-a)^2} + \sum_{m=1}^{N_1}\frac{S_m'(0)}{a_m - a}.
\end{equation}
Let $a\to a_l$ for an $l$. At the vicinity of $a_l$ we have $\ds S(a) = \frac{S_l(a-a_l)}{(a-a_l)^2}$ and therefore
$$
\lim_{a\to a_l} \left( S(a) - \frac{S_l'(0)}{(a-a_l)} - \frac{S_l(0)}{(a-a_l)^2}\right) = \frac{1}{2}S_l''(0).
$$
Obtained identities yield the expressions for $\F_{2,0}$ and $\F_{0,2}$ \Ref{PrepSUxSU}. 

In these computations the exact form of the residue function was immaterial, this proof suits for an arbitrary (non-conformal) theory.
 
In order to complete proof of the two-instanton consistency we use the special form of the residue functions given by \Ref{STdef}. In fact, the only property we are going to use is the following
$$
\begin{aligned}
\frac{\pd S_l^\pm (0)}{\pd b_p} &= \frac{S_l^\pm (0)}{b_p \pm a_l \pm M}, & \frac{\pd T_p^\pm(0)}{\pd a_l} &= \frac{T^\pm_p(0)}{a_l \pm b_p + M}.
\end{aligned}
$$
Combining it with \Ref{S/(x-a)} we observe that for the bifundamental of types $(N_1, \cc{N}_2)$ (``$-$'') and $(N_1,N_2)$ (``$+$'') there is a perfect agreement of all results.


\subsection{Concluding remarks}

We have discussed a method which allows us to extract the Seiberg-Witten curves from the formal expression for the prepotential, provided by instanton counting. The method uses some symmetry which present in cubic curves. Our logic was to search a solution of a product equation in an artificially restricted set of functions. This strategy was inspired by the fact, that in all cases, when cubic curve is known, it does possess such a symmetry. However, it can be justified only by the fact, that the method works well, as it was tested explicitly up two instantons.

When the number of groups in the product is greater that two, the situation is less optimistic. For example, for the case considered by Witten in \cite{PrepFromM} ($\SU(N_1)\times \SU(N_2)\times \dots \times \SU(N_n)$ interacting with the help of $n-1$ massless bifundamental multiplets $(N_a,\cc{N}_{a+1})$, $a=1,\dots,n-1$). For this model instead of \Ref{FacG1G2} we have the following system of product equations:
\begin{equation}
\label{nProd}
\begin{aligned}
y^{(1)+}(t)y^{(1)-}(t) &= q_1 Q_1(t) y^{(2)}(t),& t&\in[\a_l^{(1)-},\a_l^{(1)+}] \\
y^{(2)+}(t)y^{(2)-}(t) &= q_2 Q_2(t) y^{(1)}(t)y^{(3)}(t),& t&\in[\a_l^{(2)-},\a_l^{(2)+}] \\
\cdots \\
y^{(a)+}(t)y^{(a)-}(t) &= q_a Q_a(t) y^{(a-1)}(t)y^{(a+1)}(t), & t&\in[\a_l^{(a)-},\a_l^{(a)+}], \\
\cdots \\
y^{(n)+}(t)y^{(n)-}(t) &= q_n Q_n(t) y^{(n-1)}(t), & t&\in[\a_l^{(n)-},\a_l^{(n)+}].
\end{aligned}
\end{equation}
It is not clear how to write the  transformations analogous to \Ref{G1G2Root} which could help to solve the system. However there is an argument why they \emph{must} exist. As it was mentioned in the end of Section \ref{ThermoProduct}, when we deal with the $n$ gauge group product we, at first sight, have $n$ Seiberg-Witten differential and $n$ Seiberg-Witten curve. However, as states the M-theory approach, there are cases when only one of them is independent. It suggests, that we should have a group which acts on functions $y^{(a)}(z)$ and which generalized $\Integer_3$ acting as \Ref{G1G2Root}. If we believe, that there is always only one Seiberg-Witten differential and Seiberg-Witten curve, we should be able to find such a group for all imaginable models. It would be interesting to investigate  this question.


\appendix


\section{Instanton counting}

\label{InstCount}

In this appendix we briefly recall some aspects of the instanton counting \cite{SWfromInst,ABCD} with focus on the thermodynamical limit \cite{SWandRP,SPinSW,MyThesis}. The story is not supposed to be self-consistent, therefore for details the reader is invited to consult cited articles. 


\subsection{Equivariant index}

The ADHM construction \cite{ADHM,InstAndRec} for instantons for the gauge group $\SU(N)$ whose instanton number equals $k$ is given by the following complex:
$$
\begin{CD}
\V \otimes \L^{-1} @>\t>> \V \otimes \S_- \oplus \W @>\s>>  \V \otimes \L
\end{CD}
$$
where 
$$
\begin{aligned}
\tau &= \left(
\begin{array}{c}
B_1 \\
B_2 \\
I
\end{array}
\right), & \sigma &= (B_2,-B_1,J), & &\mbox{where} & B_1,B_2:\V&\to\V & &\mbox{and} & J^\dag,I:\V&\to\W.
\end{aligned}
$$
The complex property $\t\s = 0$ is insured by the ADHM equations
$$
\begin{aligned}
{[}B_1,B_2{]} + IJ &= 0, & &\Leftrightarrow, & \t\s &= 0 \\
[B_1,B_1^\dag] + [B_2,B_2^\dag] + II^\dag - J^\dag J &= 0 & &\Leftrightarrow & \t\t^\dag - \s^\dag \s &= 0.
\end{aligned}
$$
As for vector spaces we have $\V \isom \Compl^k$ is the space of the fundamental action of the dual (in sens of \cite{InstAndRec}) group, which is $\UU(k)$ in our case, $\W\isom \Compl^N$ is the space of fundamental action of the gauge group, $\S_- \isom \Compl^2$ is the space of left-handed spinors and $\L \isom \Compl$ is a fiber of the square root of the determinant bundle.

The equivariant Chern character for the universal bundle $\E$ is given by (see \cite{SmallInst,SWfromInst} for some details)
\begin{equation}
\label{ChernFund}
\begin{aligned}
\Ch_q(\E) &\equiv \Tr_\E(q) = \Tr_\W(q) + \Tr_\V(q)\Big(\Tr_{\S_-}(q)-\Tr_\L(q)-\Tr_{\L^{-1}}(q) \Big) \\
&= \sum_{l=1}^N \e^{ia_l} - (\e^{i\eps_1}-1)(\e^{i\eps_2}-1)\sum_{i=1}^k \e^{i\phi_i - i \eps_+}.
\end{aligned}
\end{equation}
Here $a_1,\dots,a_l$ are Higgs vevs, $\phi_1,\dots,\phi_k$ are parameters of the dual group maximal torus action, $\eps_1$ and $\eps_2$ are parameters of the Lorentz deformation of the theory. $\ds \eps_\pm = \frac{\eps_1\pm \eps_2}{2}$.
 
The equivariant index of the Dirac operator for the fundamental representation $N$ is given by the equivariant Atiyah-Singer theorem:
\begin{equation}
\label{IndexFund}
\begin{aligned}
\Ind_q^N &= \int_{\Compl^2} \Ch_q(\E)\Td_q(\Compl^2) = \frac{{\left.\Ch_q(\E)\right|}_{z_1=z_2=0}}{(\e^{i\eps_1}-1)(\e^{i\eps_2}-1)} \\
&= \frac{1}{(\e^{i\eps_1} -1)(\e^{i\eps_2}-1)} \sum_{l=1}^N \e^{ia_l} - \sum_{i=1}^k \e^{i\phi_i - i \eps_+}.
\end{aligned}
\end{equation}
The same argumentation is used to get the equivariant indices for other representations. For example for the $(N,\cc{N})$ which contains the adjoint one we have
\begin{equation}
\label{IndexAdjoint}
\begin{aligned}
\Ind_q^{(N,\cc{N})} &= \int_{\Compl^2} \Ch_q(\E \otimes \cc{\E})\Td_q(\Compl^2) = \frac{{\left.\Ch_q(\E)\Ch_q(\cc{\E})\right|}_{z_1=z_2=0}}{(\e^{i\eps_1}-1)(\e^{i\eps_2}-1)} = \frac{N + \sum_{l\neq m}^N \e^{ia_l-ia_m}}{(\e^{i\eps_1}-1)(\e^{i\eps_2} - 1)}\\
&- \sum_{i=1}^k\sum_{l=1}^N \left( \e^{i\phi_i - ia_l- i\eps_+} + \e^{- i\phi_i + ia_l - i\eps_+}\right) + (\e^{-i\eps_1}-1)(\e^{-i\eps_2}-1)\left( k + \sum_{i\neq j}^k \e^{i\phi_i - i\phi_j} \right).
\end{aligned}
\end{equation}
At the same way we can get equivariant indices for symmetric and antisymmetric representations. 


\subsection{Partition function}

In fact, all  formulae simplify when we put (after having performed the contour integration in \Ref{PartitionInst}) $\eps_1 = - \eps_2 = \hbar$.  As it was shown in \cite{SWfromInst,ABCD} the partition function of the theory (the expectation value of the identity operator) is given by
\begin{equation}
\label{Partition}
{\< 1 \>}_a = Z^{pert}(a,m,\gL;\eps) Z^{inst}(a,m,\gL;\eps) = \exp \frac{1}{\eps_1\eps_2} \F(a,m,\gL;\eps),
\end{equation}
where $\F(a,m,\gL;\eps) = \F(a,m,\gL) + \hbar^2 \F^{(1)}(a,m,\gL) + \hbar^4 \F^{(2)}(a,m,\gL)+ \dots$. Here $\F(a,m,\gL)$ is the Seiberg-Witten prepotential which defined the low-energy effective Wilsonian action, and $\F^{(g)}(a,m,\gL)$, $g=1,2,\dots$ are higher genius corrections. $Z^{pert}(a,m,\gL;\eps)$ is the perturbative contribution to the partition function, and
\begin{equation}
\label{PartitionInst}
Z^{inst}(a,m,\gL;\eps) = 1 + \sum_{k=1}^\infty q^k \oint \prod_{i=1}^k\frac{\dd \phi_i}{2\pi i} z_k(a,m,\phi;\eps),
\end{equation}
where $\ds q = \e^{2\pi i \t(\gL)} = \gL^\b \e^{2\pi i \t_0}$ and $\ds \t(\gL) = \t_0 + \frac{\b}{2 \pi i} \ln \gL$ is the running under the RG-flow complex coupling constant.. The integrand $z_k(a,m,\phi;\eps)$ is related to the equivariant index of the Dirac operator \Ref{IndexFund} via the following transformation ($n_\a = \pm 1$)
\begin{equation}
\label{zFromInd}
\Ind_q = \sum_\a n_\a \e^{i w_\a} \mapsto z_k(a,m,\phi;\eps) = \prod_\a {\left(w_\a\right)}^{n_\a}.
\end{equation}

The first summand in \Ref{IndexFund} under this transformation becomes an infinite product and therefore requires a regularization. It is independent of $\phi_i$s and determines the perturbative contribution. Any term of form $\ds \frac{\e^{ix}}{(\e^{i\eps_1}-1)(\e^{i\eps_2}-1)} $ contributes to $Z^{pert}(a,m,\gL;\eps)$ as
$$
\exp  \frac{1}{\eps_1\eps_2} \Big(\k_\gL(x) + O(\hbar^2)\Big),
$$
where 
$$
\k_\gL(x) = \frac{1}{2}x^2\left(\ln \left|\frac{x}{\gL}\right|-\frac{3}{2} \right).
$$
As an example we can compute the perturbative contribution due to the gauge field:
$$
\eps_1\eps_2 \ln Z^{pert}(a,\gL;\eps) = \ds\sum_{l\neq m}^N \k_\gL (a_l - a_m) + O(\hbar^2) =  \ds\sum_{l\neq m}^N \k (a_l - a_m) - N \ln \gL \sum_{l=1}^N a_l^2 + O(\hbar^2),
$$
where $\ds \k(x) = \k_1(x) = \frac{1}{2}x^2 \left( \ln |x| - \frac{3}{2}\right)$. In general case we have, up to an immaterial constant (a function of masses)
\begin{equation}
\label{PertGauge}
Z^{pert}(a,\gL;\eps) = \exp\frac{1}{\eps_1\eps_2} \Big(\F^{pert}(a,\gL) + O(\hbar^2) \Big) = \exp\frac{1}{\eps_1\eps_2} \left(\F^{pert}(a,1) - \frac{\b}{2} \ln \gL\sum_{l=1}^N a_l^2 + O(\hbar^2) \right),
\end{equation}
where $\F^{pert}(a,\gL)$ is the perturbative prepotential.

Other summands can be used to find $z_k(a,m,\phi;\eps)$. For example for the fundamental representation we have
\begin{equation}
\label{zFund}
z_k^N(a,m,\phi;\eps) = \prod_{i=1}^k (\phi_i +m - \eps_+).
\end{equation}
In that follows it is convenient to redefine $m - \eps_+ \mapsto m$. 


\subsection{Thermodynamical limit}
\label{Thermo}

In the removing Lorentz deformation limit ($\eps_1,\eps_2\to 0$) the sum \Ref{PartitionInst} is dominated by a single term with $\ds k \sim \frac{1}{\eps_1\eps_2} \to \infty$. When $k$ is large the integration in each summand in \Ref{PartitionInst} can be replace by the functional integration over the $\phi_i$'s density $\ds\r(x) = \eps_1\eps_2 \sum_{i=1}^k \d(x-\phi_i)$, which is normalized in such a way that its integral remains finite in the limit $k\to\infty$. In order to include also the perturbative contributions \Ref{PertGauge} it is convenient to introduce the profile function (in \cite{SWandRP} this function was associated with the shape of random partitions)
\begin{equation}
\label{Profile}
f(x) = \sum_{l=1}^N |x-a_l| - 2\eps_1\eps_2 \sum_{i=1}^k \d(x-\phi_k) = \sum_{l=1}^N |x-a_l| - 2 \r(x).
\end{equation}
This definition allows us to establish the following properties of the profile function:
\begin{equation}
\label{profileProp}
\begin{aligned}
\frac{1}{2}\int_\Real \dd x f''(x) &= N, & \frac{1}{2}\int_\Real \dd x f''(x) x &= \sum_{l=1}^N a_l = 0, & \frac{1}{2}\int_\Real \dd x f''(x) x^2 &= \sum_{l=1}^N a_l^2 -2\eps_1\eps_2 k.
\end{aligned}
\end{equation}
Note that the last expression together with \Ref{PertGauge} allow us to recast each summand in \Ref{Partition} in the leading order of $\hbar^2$ as follows
\begin{multline}
\label{Lambda}
Z^{pert}(a,m,\gL;\eps) q^k = Z^{pert}(a,m,\gL;\eps) \gL^{\b k} \e^{2\pi i k \t_0} \\
= \exp \frac{1}{\eps_1\eps_2} \left( \F^{class}(a,m) + \F^{pert}(a,m,1) - \frac{\pi i}{2}\t(\gL) \int \dd x f''(x) x^2 + O(\hbar^2) \right),
\end{multline}
where $\F^{class}(a,m)$ is the classical prepotential, in our example it is $\ds \F^{class}(a) = \pi i \t_0 \sum_{l=1}^N a_l^2$. Remarkably this expression contains not only the perturbative contribution to the prepotential, but also the classical one.

In the thermodynamical limit we have (note that the perturbative corrections are also included)
$$
Z(a,m,\gL;\eps) \sim \int \D f \exp\left\{-\frac{1}{\eps_1\eps_2} \Big(H[f] + O(\hbar^2) \Big)\right\}
$$
where $H[f]$ is the Hamiltonian which can by obtained directly from the equivariant index \Ref{IndexFund} as follows
$$
\Ind_q = \sum_\a n_\a \e^{i w_\a} \mapsto H = - \eps_1\eps_2\lim_{\eps_1\eps_2\to 0} \sum_\a n_\a \ln \left| w_\a\right|.
$$

For other classical gauge group ($\Sp(N)$ and $\SO(2n + \chi)$ where $\chi = 0$ or $1$) we define the profile functions as follows (for the symplectic case we represent the instanton number as follows: $\ds k_{\Sp} = 2r + \xi$, where $r$ is integer and $\xi = 0$ or $1$)
\begin{equation}
\label{ProfSOSp}
\begin{aligned}
f_{\SO(2n+\chi)}(x) &= \sum_{l=1}^n \Big(|x-a_l| + |x+a_l| \Big) + \chi |x| - 2\eps_1\eps_2 \sum_{i=1}^{k_{\SO}}\Big(\d(x-\phi_i) + \d(x+\phi_i) \Big), \\
f_{\Sp(N)}(x) &= \sum_{l=1}^N \Big( |x-a_l| + |x+a_l| \Big) -2\eps_1\eps_2 \left[ \sum_{j=1}^r \Big(\d(x-\vf_j) + \d(x+\vf_j) \Big)+ \xi \d(x)\right].
\end{aligned}
\end{equation}
These functions are symmetric: $f(x) = f(-x)$. With the help of the profile functions \Ref{Profile} and  \Ref{ProfSOSp} we obtain the Hamiltonians for numerous  models \cite{SPinSW,MyThesis} (see Table \ref{Hams}, also we have put there some expressions for the bifundamental representations obtained in Section \ref{ThermoProduct}). Note that \Ref{Lambda} shows that all $\gL$-dependence can be localized in the following term of the total Hamiltonian:
$$
+ \frac{\pi i}{2} \t(\gL)  \int \dd x f''(x) x^2 = - \pi i\t(\gL) \int \dd x f'(x) x.
$$ 

\begin{table}
\begin{center}
\begin{tabular}{||c|c||c||}
\hhline{|t:=:=:t:=:t|}
\textbf{Group} & \textbf{Multiplet} & \textbf{Contribution to  $\mbf{H[f]}$}\\
\hhline{|:=:=::=:|}	 
& Adjoint, gauge & $\ds -\frac{1}{4}\int \dd x\dd y f''(x)f''(y)\k(x-y)$\\
\hhline{||~|-||-||}
& Fundamental & $\ds  \frac{1}{2} \int \dd x f''(x) \k(x+m)$\\
\hhline{||~|-||-||}
$\SU(N)$ & Symmetric & $\ds \frac{1}{8}\int \dd x\dd y f''(x)f''(y) \k(x+y + m) +  \int \dd x f''(x) \k(x + m/2)$ \\
\hhline{||~|-||-||}
& Antisymmetric & $\ds \frac{1}{8} \int \dd x\dd y f''(x)f''(y)\k(x + y + m) -  \int \dd x f''(x) \k(x + m/2)$\\
\hhline{||~|-||-||}
& Adjoint, matter & $\ds \frac{1}{4} \int \dd x\dd y f''(x)f''(y)\k(x - y + m)$\\
\hhline{|:=:=::=:|}
& Adjoint, gauge & $\ds-\frac{1}{8} \int \dd x\dd y f''(x)f''(y)\k(x + y) + \int \dd x f''(x) \k(x)$\\
\hhline{||~|-||-||}
$\SO(N)$ & Fundamental & $\ds \frac{1}{2} \int \dd x f''(x) \k(x+m)$\\
\hhline{||~|-||-||}
& Adjoint, matter & $\ds \frac{1}{8} \int \dd x\dd y f''(x)f''(y)\k(x + y + m) -  \int \dd x f''(x) \k(x + m/2)$ \\
\hhline{|:=:=::=:|}
& Adjoint, gauge & $\ds -\frac{1}{8} \int \dd x\dd y f''(x)f''(y)\k(x + y) - \int \dd x f''(x)\k(x)$\\
\hhline{||~|-||-||}
$\Sp(N)$ & Fundamental & $\ds \frac{1}{2} \int \dd x f''(x) \k(x+m)$\\
\hhline{||~|-||-||}
& Antisymmetric & $\ds\frac{1}{8} \int \dd x\dd y f''(x)f''(y)\k(x + y + m)  - \int \dd x f''(x) \k(x + m/2)$\\
\hhline{||~|-||-||}
& Adjoint, matter & $\ds\frac{1}{8}\int \dd x\dd y f''(x)f''(y) \k(x+y + m)  + \int \dd x f''(x)\k(x+m/2)$\\ 
\hhline{|:=:=::=:|}
$\SU(N_1)$& Bifund $(N_1,N_2)$ & $\ds \frac{1}{4}\int \dd x\dd y f_1''(x)\k(x +  y + m) f_2''(y)$\\
\hhline{||~|-||-||}
$\times \SU(N_2)$ & Bifund $(N_1,\cc{N}_2)$ & $\ds \frac{1}{4}\int \dd x\dd y f_1''(x)\k(x -  y + m) f_2''(y)$\\
\hhline{|b:=:=:b:=:b|}
\end{tabular}
\end{center}
\caption{Hamiltonians}\label{Hams}
\end{table}

The prepotential is defined by the minimizer  $f_\star(x)$ of the corresponding Hamiltonian. Its supporter is the union of disjoint intervals $\ds\supp f_\star(x) = \bigcup_{l=1}^N \g_l$ where $\g_l = [\a_l^-,\a_l^+]$, $l=1,\dots,N$. Each interval $\g_l$ contains a single Higgs expectation value, $a_l \in \g_l$. This fact shows that in spite of the quadratic form of the Hamiltonians, the minimizer is defined as a solution of highly non-linear integral equation
\begin{equation}
\label{SP}
\begin{aligned}
\frac{1}{\pi i} \frac{\d H[f]}{\d f'(t)} &= \xi_l +  t \t(\gL) = \xi_l +  t \t_0  + t \frac{1}{2\pi i}\ln \gL^\b  , & t&\in \g_l,
\end{aligned}
\end{equation}
where $\xi_l$ are certain constants, which, in general, can be different for different cuts. When theory is not conformal ($\b>0$) we can neglect the second term in favor of the third. However for the conformal theories it becomes important. 

The derivative of this equation can be seen as a difference equation for the primitive of the resolvent of the profile function
\begin{equation}
\label{FandPhiDef}
F(z) = \frac{1}{4\pi i} \int \dd x f''(x) \ln(z-x).
\end{equation}
This function is supposed to be piecewise holomorphic at $\Compl \setminus \g$, where $\ds \g = \bigcup_{l=1}^N \g_l$. The piecewise holomophicity means holomorphicity in any compact disjoint to $\g$ and that the function $F(z)$ behaves integrably when $z$ gets close to $\g$.
 
Indeed, consider an $\SU(N)$ model with $N_f < 2N$ fundamental matter. Then the derivative of \Ref{SP} gets the following form:
$$
\begin{aligned}
\frac{1}{2\pi i} \int \dd x f''(x)\ln |t-x| - \frac{1}{2\pi i}\sum_{f=1}^{N_f} \ln |t+m_f| &= \t, & t&\in \g_l.
\end{aligned}
$$
With the help of \Ref{FandPhiDef} it can be rewritten as follows:
\begin{equation}
\label{DifferenceExample}
\begin{aligned}
F^+(t) + F^-(t) - \frac{1}{2\pi i} \sum_{f=1}^{N_f} \ln |t+m_f| &= \t, & t&\in \g_l,
\end{aligned}
\end{equation}
where by $F^+(t)$ and $F^-(t)$ we denote the values of the function at the upper and lower side of cuts $\g_l$. Based on the Table \ref{Hams} we can construct the table of the contribution to the difference equation \Ref{DifferenceExample} (Table \ref{Diff}).

\begin{table}
\begin{center}
\begin{tabular}{||c|c||c||}
\hhline{|t:=:=:t:=:t|}
\textbf{Group} & \textbf{Multiplet} & \textbf{Contribution to the  Difference Equation}\\
\hhline{|:=:=::=:|}	 
& Adjoint, gauge & $\ds F^+(t) + F^-(t)$\\
\hhline{||~|-||-||}
& Fundamental & $\ds  -\frac{1}{2\pi i} \ln |t + m|$\\
\hhline{||~|-||-||}
$\SU(N)$ & Symmetric & $\ds -F(-t-m) - \frac{1}{\pi i} \ln\left|t+ \frac{m}{2}\right|$ \\
\hhline{||~|-||-||}
& Antisymmetric & $\ds -F(-t-m) + \frac{1}{\pi i} \ln \left|t+\frac{m}{2}\right|$\\
\hhline{||~|-||-||}
& Adjoint, matter & $\ds -2F(t+m)$\\
\hhline{|:=:=::=:|}
& Adjoint, gauge & $\ds \frac{1}{2} F^+(t) + \frac{1}{2}F^-(t) - \frac{1}{\pi i} \ln |t|$\\
\hhline{||~|-||-||}
$\SO(N)$ & Fundamental & $\ds - \frac{1}{4\pi i} \ln |t+m| - \frac{1}{4\pi i} \ln |-t+m|$\\
\hhline{||~|-||-||}
& Adjoint, matter & $\ds - \frac{1}{2}F(t+m) -\frac{1}{2} F(-t+m) + \frac{1}{2\pi i} \ln \left| t + \frac{m}{2}\right| + \frac{1}{2\pi i} \ln \left|-t+\frac{m}{2} \right|$ \\
\hhline{|:=:=::=:|}
& Adjoint, gauge & $\ds \frac{1}{2}F^+(t) + \frac{1}{2}F^-(t) + \frac{1}{\pi i} \ln |t|$\\
\hhline{||~|-||-||}
$\Sp(N)$ & Fundamental & $\ds -\frac{1}{4\pi i} \ln |t+m| - \frac{1}{4\pi i} \ln |-t+m|$\\
\hhline{||~|-||-||}
& Antisymmetric & $\ds - \frac{1}{2} F(t+m) - \frac{1}{2} F(-t+m) + \frac{1}{2\pi i} \ln \left| t + \frac{m}{2}\right| + \frac{1}{2\pi i} \ln \left| -t + \frac{m}{2}\right|$\\
\hhline{||~|-||-||}
& Adjoint, matter & $\ds - \frac{1}{2} F(t+m) - \frac{1}{2} F(-t+m) + \frac{1}{2\pi i} \ln \left| t + \frac{m}{2}\right| + \frac{1}{2\pi i} \ln \left| -t + \frac{m}{2}\right|$\\ 
\hhline{|:=:=::=:|}
$\SU(N_1)$& Bifund $(N_1,N_2)$ & $\ds -F_2 (-t-m)$ for $f_1(x)$ and  $\ds -F_1(-t-m)$ for $f_2(x)$\\
\hhline{||~|-||-||}
$\times \SU(N_2)$ & Bifund $(N_1,\cc{N}_2)$ & $\ds -F_2(t + m)$ for $f_1(x)$ and $\ds -F_2(t-m)$ for $f_2(x)$\\
\hhline{|b:=:=:b:=:b|}
\end{tabular}
\end{center}
\caption{Contribution to the Difference equation}\label{Diff}
\end{table}

On the complex plane for $z$ the supporter of the minimizer becomes a set of cuts. We can introduce the $A$-cycles, such that $A_l$ surrounds $\g_l$ and a complementary set of dual $B$-cycles $B_l$ such that $A_l \# B_m = \d_{l,m}$. Then the Seiberg-Witten prepotential $\F(a,m,\gL)$ can be computed as follows
\begin{equation}
\label{A-cycles}
\begin{aligned}
\oint_{A_l} \l(z) \dd z &= a_l, & 2\pi i\oint_{B_l} \l(z) \dd z &= \frac{\pd \F}{\pd a_l} = a^l_D.
\end{aligned}
\end{equation}
where $\l(z) = z F'(z) \dd z$ can be associated with the Seiberg-Witten differential. Usually one introduces another function, the exponent of $F(z)$:
\begin{equation}
\label{yDef}
y(z) = \exp 2\pi i F(z).
\end{equation}
Then the Seiberg-Witten differential takes the familiar form $\ds\l(z) = \frac{1}{2\pi i}z \frac{\dd y(z)}{y(z)}$.


\section{About double ramification points}
\label{doubleRamification}

In this Appendix we justify the statement done after the formula \Ref{Fassym}, which says that $y(z)$ has only double ramification points in the endpoints of $\g_l \ni a_l$.

Recall some facts about the difference equations. If a function $\Phi(z)$ which decays at infinity as $z^{-1}$ satisfies the following difference equation on the contour $\g = [a,b]$:
\begin{equation}
\label{SokhotskiPlemeli}
\begin{aligned}
\Phi^+(t) - \Phi^-(t) &= \phi(t), & t&\in \g,
\end{aligned}
\end{equation}
then if this function is piecewise holomorphic it is unique and given by the Sokhotski-Plemelj formula
$$
\Phi(z) = \frac{1}{2\pi i} \int_\g \dd t \frac{\phi(t)}{t-z}.
$$
Suppose now that $\Phi(z)$ satisfies another difference equation: $\Phi^+(t) + \Phi^-(t) = 0$ when $t\in \g$. The logarithm of this function satisfies the difference equation \Ref{SokhotskiPlemeli} with $\phi(t) = \ln(-1) = \pi i + 2\pi i k$, where $k \in \Integer$. Then the Sokhotski-Plemelj formula shows that up to a multiplicative constant we have
$$
\Phi_0(z) = {\left(\frac{a-z}{b-z}\right)}^{k + \frac{1}{2}}.
$$  
Therefore only double ramification points are admitted. If $\Phi(z)$ satisfies an inhomogeneous equation, such as $\Phi^+(t) + \Phi^-(t) = \phi(t)$, we take at the vicinity of the contour $\Phi(z) = \Phi_0(z) + \phi(z)/2$, thus we find only double ramification points as well.

More rigorously we can proceed as follows: note that at $\g$ we have 
$$
\begin{aligned}
\frac{\Phi^+_0(t)}{\Phi^-_0(t)} &= -1, & t &\in \g.
\end{aligned}
$$
Therefore the solution of the inhomogenious equation satisfies
$$
\frac{\Phi^+(t)}{\Phi^+_0(t)} - \frac{\Phi^-(t)}{\Phi^-_0(t)} = \frac{\phi(t)}{\Phi^+_0(t)}.
$$
Applying the Sokhotsky-Plemelj formula we get finally
$$
\Phi(z) = \frac{\Phi_0(z)}{2\pi i} \int_\g \dd t \frac{\phi(t)}{\Phi_0^+(t)(t-z)} + P(z)\Phi_0(z),
$$
where $P(z)$ is a polynomial, whose form is determined by the boundary conditions at infinity. It is clear that only the double ramification points appear in $\Phi(z)$. The conclusion holds even if $\phi(z)$ depends functionally of $\Phi(z)$. 

For more details on the difference equations theory see, for example \cite{Muskhelishvili,Litvinchuk}.
 

\providecommand{\bysame}{\leavevmode\hbox to3em{\hrulefill}\thinspace}
\providecommand{\href}[2]{#2}


\begin{thebibliography}{10}

\bibitem{Curves}
E.~Arbarello, M.~Cornalba, P.A. Griffiths, and J.~Harris, \emph{Geometry of
  algebraic curves, vol. {I}}, Springer-Verlag, New-York, Berlin, Heidelberg,
  Tokyo, 1985.

\bibitem{ADHM}
M.~F. Atiyah, N.~J. Hitchin, V.~G. Drinfeld, and Yu.~I. Manin,
  \emph{Construction of instantons}, Phys.Lett. \textbf{65A} (1978), 185.

\bibitem{RGMatone}
Giulio Bonelli and Marco Matone, \emph{Nonperturbative renormalization group
  equation and beta function in {$\N=2$} {SUSY} {Y}ang-{M}ills}, Phys. Rev.
  Lett. \textbf{76} (1996), 4107--4110,
  \href{http://arxiv.org/abs/hep-th/9602174}{\tt arXiv:hep-th/9602174}.

\bibitem{MtheoryAndSW-SOandSp}
A.~Brandhuber, J.~Sonnenschein, S.~Theisen, and S.~Yankielowicz,
  \emph{{M}-theory and {S}eiberg-{W}itten curves: {O}rthogonal and {S}ymplectic
  groups}, Nucl.Phys. \textbf{B504} (1997), 175--188,
  \href{http://arxiv.org/abs/hep-th/9705232}{\tt arXiv:hep-th/9705232}.

\bibitem{RecursionEqn}
Gordon Chan and Eric D'Hoker, \emph{Instanton recursion relations for the
  effective prepotential in {$\N = 2$} super {Y}ang-{M}ills}, Nucl. Phys.
  \textbf{B564} (2000), 503--516,
  \href{http://arxiv.org/abs/hep-th/9906193}{\tt arXiv:hep-th/9906193}.

\bibitem{InstAndRec}
E.~Corrigan and P.~Goddard, \emph{Construction of instantons and monopole
  solutions and reciprocity}, Annals Phys. \textbf{154} (1984), 253.

\bibitem{HKP-SU}
E.~D'Hoker, I.~M. Krichever, and D.~H. Phong, \emph{The effective prepotential
  of {$\mathcal{N}=2$} supersymmetric {$SU(N_c)$} gauge theories},
  \href{http://arxiv.org/abs/hep-th/9609041}{\tt arXiv:hep-th/9609041}.

\bibitem{RGequation}
Eric D'Hoker, I.~M. Krichever, and D.~H. Phong, \emph{The renormalization group
  equation in {$\N = 2$} supersymmetric gauge theories}, Nucl. Phys.
  \textbf{B494} (1997), 89--104, \href{http://arxiv.org/abs/hep-th/9610156}{\tt
  arXiv:hep-th/9610156}.

\bibitem{JoseMarcosMarta}
Jose~D. Edelstein, Marta Gomez-Reino, Marcos Marino, and Javier Mas,
  \emph{{$\mathcal{N} = 2$} supersymmetric gauge theories with massive
  hypermultiplets and the {W}hitham hierarchy}, Nucl. Phys. \textbf{B574}
  (2000), 587--619, \href{http://arxiv.org/abs/hep-th/9911115}{\tt
  arXiv:hep-th/9911115}.

\bibitem{JoseMarcos}
Jose~D. Edelstein, Marcos Marino, and Javier Mas, \emph{Whitham hierarchies,
  instanton corrections and soft supersymmetry breaking in {$\mathcal{N} = 2$}
  {$\SU(N)$} super {Y}ang-{M}ills theory}, Nucl. Phys. \textbf{B541} (1999),
  671--697, \href{http://arxiv.org/abs/hep-th/9805172}{\tt
  arXiv:hep-th/9805172}.

\bibitem{PrepotentialEquation}
Tohru Eguchi and Sung-Kil Yang, \emph{Prepotentials of {$\N=2$} supersymmetric
  gauge theories and soliton equations}, Mod. Phys. Lett. \textbf{A11} (1996),
  131--138, \href{http://arxiv.org/abs/hep-th/9510183}{\tt
  arXiv:hep-th/9510183}.

\bibitem{MTheoryTested}
I.~Ennes, C.~Lozano, S.~Naculich, H.~Rhedin, and H.~Schnitzer, \emph{{M}-theory
  tested by {$\mathcal{N}=2$} {S}eiberg-{W}itten theory},
  \href{http://arxiv.org/abs/hep-th/0006141}{\tt arXiv:hep-th/0006141}.

\bibitem{EnnesMasterFunc}
I.~Ennes, S.~Naculich, H.~Rhedin, and H.~Schnitzer, \emph{Tests of {M}-theory
  from {$\mathcal{N}=2$} {S}eiberg-{W}itten theory},
  \href{http://arxiv.org/abs/hep-th/9911022}{\tt arXiv:hep-th/9911022}.

\bibitem{EnnesSUSymAnt}
\bysame, \emph{One-instanton predictions for non-hyperelliptic curves derived
  from {M}-theory}, Nucl.Phys. \textbf{B536} (1998), 245,
  \href{http://arxiv.org/abs/hep-th/9806144}{\tt arXiv:hep-th/9806144}.

\bibitem{EnnesSUSymFund}
\bysame, \emph{One instanton predictions of a {S}eiberg-{W}itten curve from
  {M}-theory: the symmetric representation of {$\SU(N)$}}, Int.J.Mod.Phys.
  \textbf{A14} (1999), 301, \href{http://arxiv.org/abs/hep-th/9804151}{\tt
  arXiv:hep-th/9804151}.

\bibitem{EnnesSU2AntFund}
\bysame, \emph{Two antisymmetric hypermultiplets in {$\mathcal{N}=2$}
  {$\SU(N)$} gauge theory: {S}eiberg-{W}itten curve and {M}-theory
  interpretation}, Nucl. Phys. \textbf{B558} (1999), 41,
  \href{http://arxiv.org/abs/hep-th/9904078}{\tt arXiv:hep-th/9904078}.

\bibitem{1instProduct}
Isabel~P. Ennes, Stephen~G. Naculich, Henric Rhedin, and Howard~J. Schnitzer,
  \emph{One-instanton predictions of {S}eiberg-{W}itten curves for product
  groups}, Phys. Lett. \textbf{B452} (1999), 260--264,
  \href{http://arxiv.org/abs/hep-th/9901124}{\tt arXiv:hep-th/9901124}.

\bibitem{GroupProductFromBranes}
Joshua Erlich, Asad Naqvi, and Lisa Randall, \emph{The {C}oulomb branch of
  $\mathcal{N} = 2$ supersymmetric product group theories from branes},
  Phys.Rev. \textbf{D58} (1998), 046002,
  \href{http://arxiv.org/abs/hep-th/9801108}{\tt arXiv:hep-th/9801108}.

\bibitem{Quiver}
Francesco Fucito, Jose~F. Morales, and Rubik Poghossian, \emph{Instantons on
  quivers and orientifolds}, JHEP \textbf{10} (2004), 037,
  \href{http://arxiv.org/abs/hep-th/0408090}{\tt arXiv:hep-th/0408090}.

\bibitem{MartaSUxSU}
Marta Gomez-Reino, \emph{Prepotential and instanton corrections in
  {$\mathcal{N} = 2$} supersymmetric {$\SU(N_1) \times \SU(N_2)$} {Y}ang
  {M}ills theories}, JHEP \textbf{03} (2003), 043,
  \href{http://arxiv.org/abs/hep-th/0301105}{\tt arXiv:hep-th/0301105}.

\bibitem{Gorsky}
A.~Gorsky, A.~Marshakov, A.~Mironov, and A.~Morozov, \emph{{RG} equations from
  {W}hitham hierarchy}, Nucl. Phys. \textbf{B527} (1998), 690--716,
  \href{http://arxiv.org/abs/hep-th/9802007}{\tt arXiv:hep-th/9802007}.

\bibitem{NewCurvesFromBranes}
Karl Landsteiner and Esperanza Lopez, \emph{New curves from branes}, Nucl.
  Phys. \textbf{B516} (1998), 273--296,
  \href{http://arxiv.org/abs/hep-th/9708118}{\tt arXiv:hep-th/9708118}.

\bibitem{N=2BranesOrientifolds}
Karl Landsteiner, Esperanza Lopez, and David~A. Lowe, \emph{{$\N = 2$}
  supersymmetric gauge theories, branes and orientifolds}, Nucl. Phys.
  \textbf{B507} (1997), 197--226,
  \href{http://arxiv.org/abs/hep-th/9705199}{\tt arXiv:hep-th/9705199}.

\bibitem{Litvinchuk}
Georgii~S. Litvinchuk, \emph{Solvability theory of boundary value problems and
  singular integral equations with shift}, {K}luwer {A}cademic {P}ublishers,
  Dordrecht, 2000.

\bibitem{SmallInst}
A.~Losev, A.~Marshakov, and N.~Nekrasov, \emph{Small instantons, little strings
  and free fermions}, \href{http://arxiv.org/abs/hep-th/0302191}{\tt
  arXiv:hep-th/0302191}.

\bibitem{WyllMar}
Marcos Mari{\~n}o and Niclas Wyllard, \emph{A note on instanton counting for
  {$\mathcal{N} = 2$} gauge theories with classical gauge groups}, JHEP
  \textbf{05} (2004), 021, \href{http://arxiv.org/abs/hep-th/0404125}{\tt
  arXiv:hep-th/0404125}.

\bibitem{MarcoRecursion}
Marco Matone, \emph{Instantons and recursion relations in {$\N=2$} {SUSY} gauge
  theory}, Phys. Lett. \textbf{B357} (1995), 342--348,
  \href{http://arxiv.org/abs/hep-th/9506102}{\tt arXiv:hep-th/9506102}.

\bibitem{Muskhelishvili}
N.I. Muskhelishvili, \emph{Singular integral equations}, Groningen, Noordholf,
  1953.

\bibitem{NaculichSUAntFund}
S.~Naculich, H.~Rhedin, and H.~Schnitzer, \emph{One-instanton test of a
  {S}eiberg-{W}itten curve from {M}-theory: the antisymmetric representation of
  {$\SU(N)$}}, Nucl. Phys. \textbf{B533} (1998), 275,
  \href{http://arxiv.org/abs/hep-th/9804105}{\tt arXiv:hep-th/9804105}.

\bibitem{SWfromInst}
N.~Nekrasov, \emph{{S}eiberg-{W}itten prepotential from instanton counting},
  Adv.Theor.Math.Phys. \textbf{7} (2004), 831--864,
  \href{http://arxiv.org/abs/hep-th/0206161}{\tt arXiv:hep-th/0206161}.

\bibitem{SWandRP}
N.~Nekrasov and A.~Okounkov, \emph{{S}eiberg-{W}itten theory and random
  partitions}, \href{http://arxiv.org/abs/hep-th/0306238}{\tt
  arXiv:hep-th/0306238}.

\bibitem{ABCD}
N.~Nekrasov and S.~Shadchin, \emph{{ABCD} of instantons}, Comm.Math.Phys.
  \textbf{253} (2004), 359--391, \href{http://arxiv.org/abs/hep-th/0404225}{\tt
  arXiv:hep-th/0404225}.

\bibitem{Prepotential}
N.~Seiberg, \emph{Supersymmetry and nonperturbative beta functions}, Phys.
  Lett. \textbf{206B} (1988), 75.

\bibitem{SeibergWitten}
N.~Seiberg and E.~Witten, \emph{Electric-magnetic duality, monopole
  condensation, and confinement in $\mathcal{N} = 2$ supersymmetric
  {Y}ang-{M}ills theory}, Nucl. Phys. \textbf{B426} (1994), 19,
  \href{http://arxiv.org/abs/hep-th/9407087}{\tt arXiv:hep-th/9407087}.

\bibitem{SeibergWittenII}
\bysame, \emph{Monopoles, duality and chiral symmetry breaking in
  $\mathcal{N}=2$ supersymmetric {QCD}}, Nucl.Phys. \textbf{B431} (1994),
  484--550, \href{http://arxiv.org/abs/hep-th/9408099}{\tt
  arXiv:hep-th/9408099}.

\bibitem{MyThesis}
Sergey Shadchin, \emph{On certain aspects of string theory/gauge theory
  correspondence}, \href{http://arxiv.org/abs/hep-th/0502180}{\tt
  arXiv:hep-th/0502180}, PhD thesis.

\bibitem{SPinSW}
\bysame, \emph{Saddle point equations in {S}eiberg-{W}itten theory}, JHEP
  \textbf{0410} (2004), 033, \href{http://arxiv.org/abs/hep-th/0408066}{\tt
  arXiv:hep-th/0408066}.

\bibitem{prepotRelation}
J.~Sonnenschein, S.~Theisen, and S.~Yankielowicz, \emph{On the relation between
  the holomorphic prepotential and the quantum moduli in {SUSY} gauge
  theories}, Phys. Lett. \textbf{B367} (1996), 145--150,
  \href{http://arxiv.org/abs/hep-th/9510129}{\tt arXiv:hep-th/9510129}.

\bibitem{PrepFromM}
E.~Witten, \emph{Solutions of four-dimensional field theories via {M}-theory},
  Nucl.Phys. \textbf{B500} (1997), 3--42,
  \href{http://arxiv.org/abs/hep-th/9703166}{\tt arXiv:hep-th/9703166}.

\end{thebibliography}

\end{document}